\documentclass[%
reprint,
superscriptaddress,
amsmath,amssymb,
aps,
floatfix,
longbibliography
]{revtex4-2}
\usepackage{times}
\usepackage{graphicx}% Include figure files
\usepackage{epsfig}
\usepackage{xcolor}
\usepackage{dcolumn}% Align table columns on decimal point
\usepackage{float}
\usepackage[breaklinks]{hyperref}% add hypertext capabilities
\hypersetup{colorlinks=true,linkcolor=blue,citecolor=blue,urlcolor=blue}
\begin{document}

%\preprint{}

\title{Topological  Majorana flat bands in the Kitaev model on a Bishamon-kikko lattice}

\author{Kiyu Fukui}
\email{k.fukui@aion.t.u-tokyo.ac.jp}
\affiliation{Department of Applied Physics, The University of Tokyo, Bunkyo, Tokyo 113-8656, Japan}
\affiliation{Department of Physical Sciences, Ritsumeikan University, Kusatsu, Shiga 525-8577, Japan}
 
\author{Yukitoshi Motome}
\affiliation{Department of Applied Physics, The University of Tokyo, Bunkyo, Tokyo 113-8656, Japan}

\date{\today}

\begin{abstract}
We unveil an interesting example of topological flat bands of Majorana fermions in quantum spin liquids. We study the Kitaev model on a periodically depleted honeycomb lattice, under a magnetic field within the perturbation theory. The model can be straightforwardly extended while maintaining the exact solvability, and its ground state is a quantum spin liquid as on the honeycomb lattice. As fractionalized excitations, there are \textit{unpaired} localized Majorana fermions in addition to the itinerant Majorana fermions and $\mathbb{Z}_{2}$ fluxes. We show that in the absence of the magnetic field the Majorana fermions have completely flat bands at zero energy, and by applying the magnetic field, they turn into topological flat bands with nonzero Chern number. By varying the anisotropy of the interactions and the magnitude of the magnetic field, we clarify that the system exhibits a variety of topological phases that do not appear in the original model. We emphasize that the topological flat bands that give this rich topology come from the hybridization of the Majorana flat bands and unpaired Majorana fermions, which is unique to the flat bands of fractionalized excitations in quantum spin liquids. Our findings would stimulate the exploration of a new type of Kitaev materials exhibiting rich topology from topological Majorana flat bands.
\end{abstract}

\maketitle
%\tableofcontents

\section{Introduction}\label{sec:intro}
Topological flat bands, which are nearly flat bands with nontrivial topology, provide a promising and tunable platform to realize a variety of intriguing correlated topological phenomena, such as fractional quantum anomalous Hall effect and fractional topological/Chern insulators~\cite{Bergholtz2013, Neupert2015}. Recently, it has been clarified that topological flat bands also host new fractional quantum Hall states~\cite{Sarkar2025}, giant anomalous Hall effect~\cite{Sullivan2025}, and $\eta$-pairing superconductivity~\cite{Herzog-Arbeitman2022}.
Moreover, they may play a pivotal role in two prominent areas that have garnered significant interest in recent years: twisted multilayer graphene~\cite{Wu2021, Ma2021} and kagome superconductors~\cite{Yin2022, Okamoto2022}. Although topological flat bands bring about various novel states of matter, a comprehensive understanding of the origin and consequences of the emergent topological flat bands still remains elusive, constituting a significant challenge. In addition, topological flat bands of magnons~\cite{Chisnell2015, Zou2025}, photons in circuit quantum electrodynamics~\cite{Petrescu2012}, and hard-core bosons mapped from ultracold polar molecules~\cite{Yao2013} have been proposed. These topological flat bands of emergent quasiparticles in strongly correlated systems have been less studied than those of electrons as exemplified above, and a large part of them remains unclear.

One of the most exotic quasiparticles that manifest in strongly correlated systems is the Majorana fermions in the Kitaev quantum spin liquid (QSL)~\cite{Motome2020, Nasu2023}. The celebrated Kitaev model~\cite{Kitaev2006}, which provides a rare example of QSLs in more than one dimension, hosts itinerant Majorana fermions and localized $\mathbb{Z}_2$ fluxes as fractional excitations. The Majorana fermion excitation constitutes gapless Dirac-like nodes at zero magnetic field.
Under a magnetic field, they are gapped out due to the broken time-reversal symmetry and the system becomes topologically nontrivial, exhibiting chiral edge modes and non-Abelian anyons that would be utilized for topological quantum computation~\cite{Kitaev2006, Kitaev2003, Nayak2008}. Stimulated by the proposal of its feasibility in spin-orbit coupled Mott insulators~\cite{Jackeli2009}, numerous intensive searches for the candidate materials have been carried out~\cite{Rau2016, Winter2017, Takagi2019, Motome2020, Motome2020a, Trebst2022, Kim2022, Rousochatzakis2024, Matsuda2025, Moller2025} for instance, for Na$_2$IrO$_3$~\cite{Chaloupka2010, Singh2010, Singh2012, Comin2012, Chaloupka2013, Foyevtsova2013, Sohn2013, Katukuri2014, Yamaji2014, HwanChun2015, Winter2016}, $\alpha$-Li$_2$IrO$_3$~\cite{Singh2012, Chaloupka2013, Winter2016}, and $\alpha$-RuCl$_3$~\cite{Plumb2014, Kubota2015, Winter2016, Yadav2016, Sinn2016}.  
In recent years, such exploration has been extended to new family of candidates, such as cobalt compounds~\cite{Liu2018, Sano2018, Liu2020, Kim2022}, iridium ilmenites~\cite{Haraguchi2018, Haraguchi2020, Jang2021}, and $f$-electron compounds~\cite{Jang2019, Xing2020, Jang2020, Ramanathan2021, Daum2021}. In addition to the solid state materials, realization of the model in ultracold atoms~\cite{Duan2003, Micheli2006, Manmana2013, Gorshkov2013, Fukui2022, Sun2023} and Rydberg atoms~\cite{Kalinowski2023, Nishad2023} has also been proposed.

In this paper, we investigate topological flat bands of the Majorana fermions in the Kitaev QSL, by introducing $1/6$ periodic site depletion to the honeycomb lattice. Such $1/6$-depleted honeycomb lattice, dubbed Bishamon-kikko lattice here, has been studied for electron systems and shown to host flat bands or topological flat bands~\cite{Hu2023, Chen2024, Ikegami2024}. We study Majorana band structures of the Kitaev model on the Bishamon-kikko lattice, and clarify that zero-energy completely flat bands appear at zero magnetic field.
Based on the perturbation theory with respect to the field strength, we also clarify the emergence of topological Majorana flat bands with nonzero Chern number under the magnetic field. Due to large Chern numbers of these flat bands, a variety of topological phases with large Chern number can be realized. We also clarified its distinctive mechanism: these large Chern number stems from hybridization of two types of Majorana fermions with nontrivial topology. Finally, we suggest the measurement of thermal Hall effect as a method to detect such large Chern number, by calculating thermal Hall conductivity. Our achievement would stimulate the exploration of a new type of Kitaev materials exhibiting rich topology from topological Majorana flat bands, and exploration of the correlated topological phenomena of the Majorana fermions yet to be discovered.

The organization of this paper is as follows. In Sec.~\ref{subsec:latticemodel}, we introduce our lattice and model. Then we derive the effective Hamiltonian based on the perturbation theory in Sec.~\ref{subsec:perturbation}. To characterize the topology of the system, we introduce the band and total Chern numbers and the numerical method to calculate them in Sec.~\ref{subsec:chern}, and we also introduce the thermal Hall conductivity as a physical quantity reflecting the topology in Sec.~\ref{subsec:kappa}. As a result, we show the band structures and topological phase diagram in Sec.~\ref{subsec:band} and \ref{subsec:phase}, respectively. As a way to detect the topology, we calculate the thermal Hall conductivity in Sec.~\ref{subsec:Hall}. In Sec.~\ref{sec:discussion}, we discuss the mechanism of those large Chern numbers, and relation between previous studies on the effect of isolated site vacancies. Finally, we summarize our findings in Sec.~\ref{sec:summary}.

\section{Model and method}\label{sec:model}

\begin{figure*}
    \centering
     \includegraphics[width=\linewidth,clip]{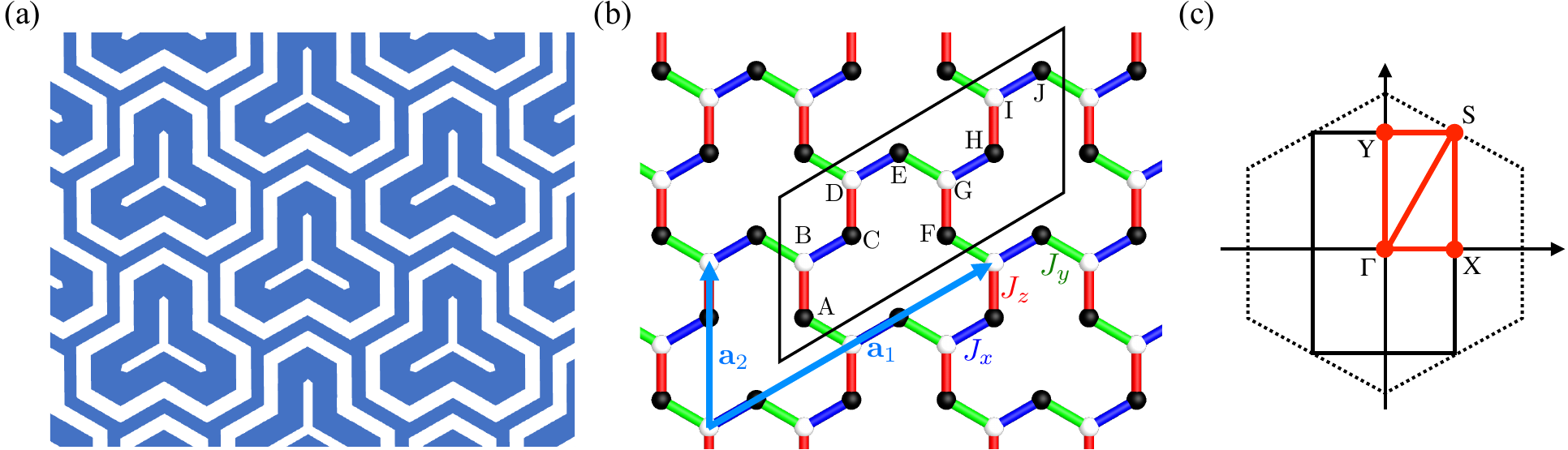}
    \caption{(a) Japanese traditional Bishamon-kikko pattern. (b) Schematic picture of the Kitaev model on a Bishamon-kikko lattice. $J_{\mu}$ represents the coupling constant of the Kitaev interactions on the $\mu$ bond ($\mu=x$, $y$, and $z$). The black parallelogram indicates the unit cell of the model contains ten lattice sites (labeled with A to J). Blue arrows represent the primitive translation vectors $\mathbf{a}_{1}$ and $\mathbf{a}_{2}$. (c) The first Brillouin zones of the model (thick line) and the original Bishamon-kikko lattice (dotted line). $\Gamma$, $\mathrm{X}$, $\mathrm{Y}$, and $\mathrm{S}$ denote the high-symmetry points and the red thick line represents the path we show the band structures along it.
}
    \label{fig:model}
\end{figure*}

\subsection{Lattice and model}\label{subsec:latticemodel}
First, we introduce the Kitaev model~\cite{Kitaev2006} on an 1/6-depleted honeycomb lattice which we call a Bishamon-kikko lattice~\cite{Ikegami2024} after the Japanese traditional Bishamon-kikko pattern, as shown in Figs.~\ref{fig:model}(a) and \ref{fig:model}(b). The lattice contains sites with coordination numbers of 2 and 3, colored in black and white in Fig.~\ref{fig:model}(b), respectively.
This lattice structure has already been realized in a van der Waals itinerant ferromagnet Fe$_{5-\delta}$GeTe$_{2}$~\cite{Wu2024}, and models of itinerant electrons on the lattice have been studied theoretically~\cite{Hu2023, Chen2024, Ikegami2024}.
Geometrically, this lattice belongs to the class of superhoneycomb systems~\cite{Shima1993}. 
Various series of periodically depleted honeycomb lattices have been explored, including those found in graphene nanostructures~\cite{Bai2010, Zhang2017}, vacancy-engineered graphenes~\cite{Souza2022, Souza2022a}, graphenes with Kekul\'e bond order~\cite{Gutierrez2016}, and photonic crystals~\cite{Ochiai2009, Lu2014}. 
%While we can consider several types of the Bishamon-kikko pattern and lattice with the plaquette shown in Fig.~\ref{fig:model}(b) depending on the way of site depletion,
Although various types of the Bishamon-kikko lattice can be constructed using the unit motif tiling shown in Fig.~\ref{fig:model}(a), we focus on the specific type shown in Figs.~\ref{fig:model}(a) and \ref{fig:model}(b) in this paper. This is because the other types have not been reported in actual materials thus far, and it is the simplest one with the minimal structural unit cell, including five sites, while we have to enlarge it to make $\pi$-flux state later. The Hamiltonian of this system is given by

\begin{equation}\label{eq:model}
\mathcal{H}_{\mathrm{K}}=-\sum_{\mu=x, y, z}\sum_{\langle i, j\rangle_{\mu}}J_{\mu}S_{i}^{\mu}S_{j}^{\mu},
\end{equation}
where the summation of $\langle i, j \rangle_{\mu}$ runs over pairs of nearest-neighbor sites $i$ and $j$ connected by $\mu$ bond, $S_{i}^{\mu}$ is the $\mu$ componenet of the $S=1/2$ spin operator at site $i$, and $J_{\mu}$ represent the coupling constant of the Kitaev interaction on the $\mu$ bond ($\mu=x$, $y$, or $z$). 
Note that our model is different from the Kitaev model on a decorated honeycomb lattice where we introduce additional sites on the center of each bond.
It can be mapped to an exactly solvable model consisting of the independent quantum spins coupled to the site-dependent Zeeman field~\cite{Mizoguchi2022, Mizoguchi2024}.

Despite $1/6$ site depletion, this model is exactly solvable through the Majorana fermion representation of the quantum spins introduced in the Kitaev's original paper~\cite{Kitaev2006}, in which spins are represented by four types of Majorana fermions ($c_{i}$, $b_{i}^{x}$, $b_{i}^{y}$, $b_{i}^{z}$) as $S_{i}^{\mu}=\mathrm{i}b_{i}^{\mu}c_{i}/2$. We set the reduced Planck constant $\hbar=1$ throughout this paper. In this representation, the spin degrees of freedom are fractionalized into itinerant Majorana fermions $c_{i}$ and the localized $\mathbb{Z}_{2}$ fluxes defined as $W_{p}=\prod_{\langle i, j\rangle_{\mu}\in p}u_{ij}$ on each plaquette labeled $p$ of the lattice; $u_{ij}^{\mu}=\mathrm{i}b_{i}^{\mu}b_{j}^{\mu}$ is a $\mathbb{Z}_{2}$ gauge variable taking $\pm1$ on the $\mu$ bond connecting site $i$ and $j$ belonging to the black and white site in Fig.~\ref{fig:model}(b), respectively. This process is pictorically shown in Fig.~\ref{fig:fractionalization}(a).
The $\mathbb{Z}_{2}$ fluxes commute with the Hamiltonian in Eq.~\eqref{eq:model}, and they are still conserved quantities even after site depletion. This enables us to group the eigenstates into different sectors of the flux configuration as in the original Kitaev model defined on a honeycomb lattice. We can find that $\pi$-flux sector, where $W_{p}=-1$ on every plaquette, gives the smallest energy. We confirm that both for the isotropic and anisotropic cases numerically in Appendix~\ref{app:flux}. While our model does not fulfill the applicability conditions, the result is consistent with Lieb's theorem, which states that the $\pi$-flux state gives the ground state in the case where the length of the plaquettes is multiples of four~\cite{Lieb1992, Lieb1994}.

As we mentioned before, here we consider an enlarged unit cell including ten sites as shown in Fig.~\ref{fig:model}(b) to make $\pi$-flux state. Applying the Majorana fermion representation, the Hamiltonian in Eq.~\eqref{eq:model} can be rewritten as
\begin{align}\label{eq:H_K}
\mathcal{H}_{\mathrm{K}}=\frac{\mathrm{i}}{4}\sum_{\mathbf{r}}\big[J_{x} (& u^{x}_{\mathrm{CB}}c_{\mathbf{r},\mathrm{C}}c_{\mathbf{r},\mathrm{B}} +  u^{x}_{\mathrm{ED}}c_{\mathbf{r},\mathrm{E}}c_{\mathbf{r},\mathrm{D}} \notag\\
+&u^{x}_{\mathrm{HG}}c_{\mathbf{r},\mathrm{H}}c_{\mathbf{r},\mathrm{G}} + u^{x}_{\mathrm{JI}}c_{\mathbf{r},\mathrm{J}}c_{\mathbf{r},\mathrm{I}})\notag\\
+ J_{y}(& u^{y}_{\mathrm{EG}}c_{\mathbf{r},\mathrm{E}}c_{\mathbf{r},\mathrm{G}} + u^{y}_{\mathrm{AD}}c_{\mathbf{r}+\mathbf{a}_{2},\mathrm{A}}c_{\mathbf{r},\mathrm{D}}\notag\\
+&u^{y}_{\mathrm{JB}}c_{\mathbf{r},\mathrm{J}}c_{\mathbf{r}+\mathbf{a}_{1},\mathrm{B}} + u^{x}_{\mathrm{FI}}c_{\mathbf{r}+\mathbf{a}_{2},\mathrm{F}}c_{\mathbf{r},\mathrm{I}})\notag\\
+J_{z}(& u^{z}_{\mathrm{AB}}c_{\mathbf{r},\mathrm{A}}c_{\mathbf{r},\mathrm{B}} + u^{z}_{\mathrm{CD}}c_{\mathbf{r},\mathrm{C}}c_{\mathbf{r},\mathrm{D}}\notag\\
+&u^{z}_{\mathrm{FG}}c_{\mathbf{r},\mathrm{F}}c_{\mathbf{r},\mathrm{G}} + u^{z}_{\mathrm{HI}}c_{\mathbf{r},\mathrm{H}}c_{\mathbf{r},\mathrm{I}})\big],
\end{align}
where $c_{\mathbf{r}, \eta}$ represents Majorana fermion at $\eta$ site in the unit cell at the position $\mathbf{r}$ ($\eta=\mathrm{A}$, $\mathrm{B}$, $\mathrm{C}$, $\mathrm{D}$, $\mathrm{E}$, $\mathrm{F}$, $\mathrm{G}$, $\mathrm{H}$, $\mathrm{I}$, or $\mathrm{J}$); $\mathbf{a}_{1}=(\sqrt{3},\ 1)$ and $\mathbf{a}_{2}=(0,\ 1)$ denote the primitive translation vectors with the unit where $\lvert \mathbf{a}_{2}\rvert=\lvert \mathbf{a}_{1}\rvert/2=1$ as shown in Fig.~\ref{fig:model}(b). 
As a result, the quantum spin model in Eq.~\eqref{eq:model} is rewritten as a nearest-neighbor tight-binding model of the $c$ Majorana fermions without interactions, as in the original honeycomb case.
In order to create the $\pi$-flux state, we have to set at least one $u_{\eta\eta'}^{\mu}$ to $-1$ while setting other $u_{\eta\eta'}^{\mu}$ to $+1$. For example, we can realize the $\pi$-flux state by setting  $u_{\mathrm{AB}}^{z}=-1$ and otherwise to $+1$. As this configuration doubles the periodicity of the system, we need to consider the enlarged unit cell containing ten sites here.

\subsection{Perturbation theory}\label{subsec:perturbation}

\begin{figure}
    \centering
     \includegraphics[width=\columnwidth,clip]{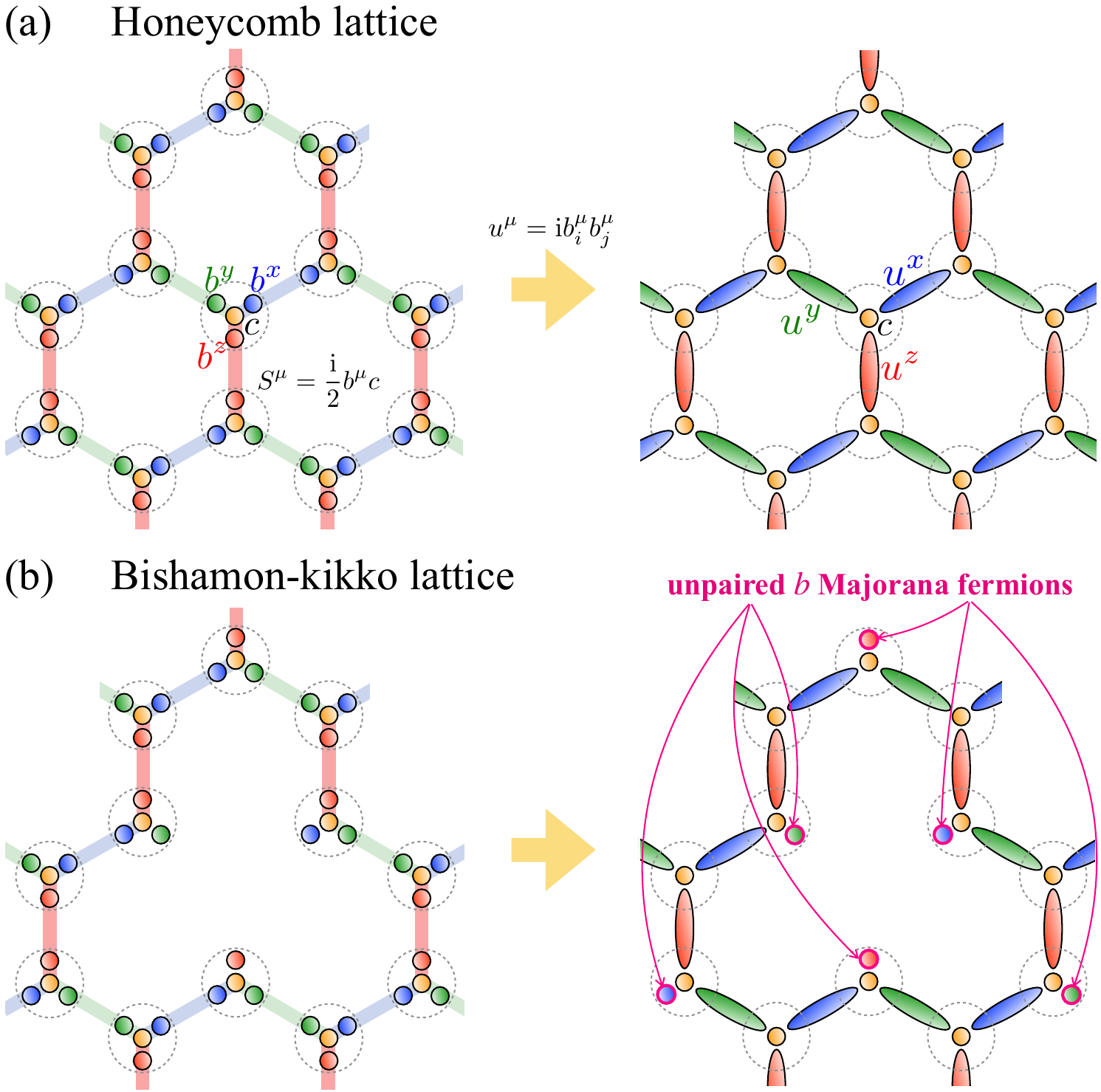}
    \caption{Schematic pictures showing the difference in the fractionalization of quantum spins between the honeycomb and Bishamon-kikko lattices. Expressing a quantum spin as $S^{\mu}=\mathrm{i}b^{\mu}c/2~(\mu=x,~y,~\text{or}~z)$ with four types of Majorana fermions $b^{x},~b^{y},~b^{z},~\text{and}~c$ (spheres in the left panels). $b$ Majorana fermions make the $\mathbb{Z}_{2}$ gauge field $u^{\mu}$ in pairs. On the Bishamon-kikko lattice, there are \textit{unpaired} $b$ Majorana fermions due to depletion (the right panels).
}
    \label{fig:fractionalization}
\end{figure}

\begin{figure}
    \centering
     \includegraphics[width=\columnwidth,clip]{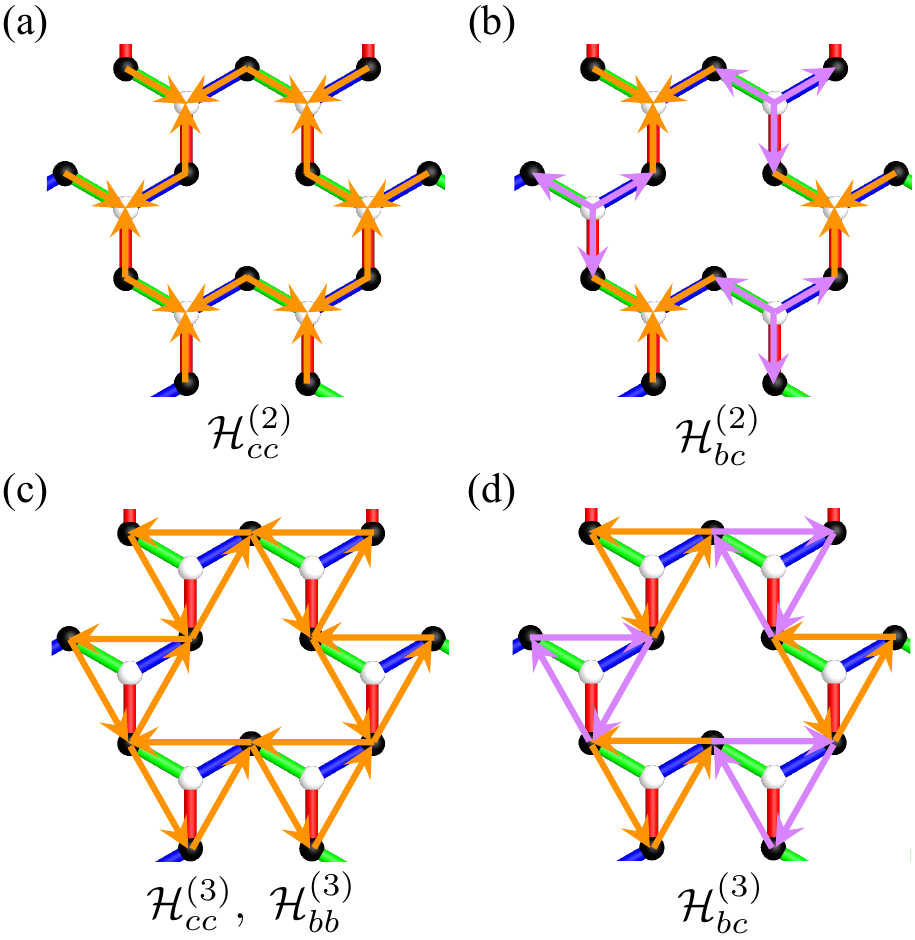}
    \caption{Schematic pictures of the hoppings of the Majorana fermions included in each term in Eq.~\eqref{eq:h_eff}, which is obtained by the third-order perturbation theory: (a) nearest-neighbor $c$-$c$, (b) nearest-neighbor $b$-$c$, (c) second-neighbor $c$-$c$ and $b$-$b$, and (d) second-neighbor $b$-$c$ hopping processes. 
Hoppings from black sites to white sites are indicated by orange arrows, and those in the opposite direction are indicated by pink arrows.
}
    \label{fig:hopping}
\end{figure}

In order to investigate the topology, we consider the effect of the magnetic field in terms of the Zeeman coupling term as
\begin{equation}
\mathcal{H}_{\mathrm{Z}}=-\sum_{i}\mathbf{h}\cdot\mathbf{S}_{i},
\end{equation}
where $\mathbf{h}=( h_{x},\ h_{y},\ h_{z})$ denotes the uniform magnetic field. By using the perturbation theory with respect to the magnetic field up to the third order~\cite{Kitaev2006}, $\mathcal{H}_{\mathrm{Z}}$ can be effectively described by additional nearest-neighbor and second-neighbor Majorana hopping terms. The resultant effective Hamiltonian is given by
\begin{equation}\label{eq:h_eff}
\mathcal{H}_{\mathrm{eff}}=\mathcal{H}_{\mathrm{K}}+\mathcal{H}_{bc}^{(1)}+\mathcal{H}_{cc}^{(2)}+\mathcal{H}_{bc}^{(2)}+\mathcal{H}_{cc}^{(3)}+\mathcal{H}_{bc}^{(3)}+\mathcal{H}_{bb}^{(3)},
\end{equation}
where the superscript of each term denotes the order of perturbation and the subscript denotes the species of Majorana fermions included in each contribution. The pictorial representation of each term is shown in Fig.~\ref{fig:hopping}, and their explicit formulas are shown in Appendix~\ref{app:perturbation}. We note that the third-order perturbation yields Majorana interactions involving products of up to ten Majorana fermion operators, which could give rise to intriguing quantum many-body effects. To focus on the topology of Majorana band structures, we here neglect those interaction terms and leave the detailed investigation for future research.

In Eq.~\eqref{eq:h_eff}, $\mathcal{H}_{\mathrm{K}}$ is unperturbed Hamiltonian expressed in the $c$ Majorana fermions in Eq.~\eqref{eq:H_K}. $\mathcal{H}_{bc}^{(1)}$, $\mathcal{H}_{bc}^{(2)}$, and $\mathcal{H}_{bc}^{(3)}$ represent on-site $b$-$c$ hybridizations, nearest-neighbor $b$-$c$ hoppings [Fig.~\ref{fig:hopping}(b)], and second-neighbor $b$-$c$ hoppings [Fig.~\ref{fig:hopping}(d)], respectively. These $b$-$c$ hybridization terms can also appear in the Kitaev model with isolated site vacancies on the honeycomb lattice (as derived in Ref.~\onlinecite{Takahashi2023}, for example). $H_{bb}^{(3)}$ denotes second-neighbor hoppings of the unpaired $b$ Majorana fermions [see Fig.~\ref{fig:hopping}(c)]. 
While this hopping is absent for isolated vacancies, it appears in the case of the Bishamon-kikkou lattice where vacancies are arranged periodically.
The remaining second- and third-order corrections, $\mathcal{H}_{cc}^{(2)}$ and $\mathcal{H}_{cc}^{(3)}$, denote the nearest-neighbor [Fig.~\ref{fig:hopping}(a)] and second-neighbor hoppings [Fig.~\ref{fig:hopping}(c)] of $c$ Majorana fermions, respectively, which also appear in the honeycomb case. The first, second, and third order perturbative corrections are of the order of $\mathcal{O}(h_{\mu})$, $\mathcal{O}(h_{\mu}^{2}/\Delta)$, and $\mathcal{O}(h^{3}_{\mu}/\Delta^{2})$, respectively. $\Delta$ represents the flux gap from the $\pi$-flux state at zero magnetic field. The flux gap is defined as an energy gap between the $\pi$-flux state and a state where a pair of fluxes on two adjacent plaquettes is excited. We estimated $\Delta$ numerically as $\Delta\simeq 0.0033$ by extrapolating the system size dependence of the excitation energy to the thermodynamic limit.

These perturbative corrections contain hopping processes involving $b$ Majorana fermions, which do not appear in the original honeycomb case in Eq.~\eqref{eq:H_K}. 
In the derivation of Eq.~\eqref{eq:H_K}, three $b$ Majorana fermions are introduced at each site, and all of them are used to construct the gauge fields $u_{ij}^\mu$ in pairs on all bonds, as the honeycomb lattice is tricoordinated [see Fig.~\ref{fig:fractionalization}(a)].
This is also true for the sites represented by the white spheres of the Bishamon-kikko lattice in Fig.~\ref{fig:model}(b).
 In contrast, as the Bishamon-kikko lattice has the lattice sites whose coordination number is $2$ [represented by black spheres in Fig.~\ref{fig:model}(b)], the effective Majorana Hamiltonian has \textit{unpaired} (or \textit{dangling}) $b$ Majorana fermions at these sites, as schematically shown in Fig.~\ref{fig:fractionalization}(b). 
 At zero magnetic field, the itinerant $c$ Majorana fermions and those unpaired $b$ Majorana fermions are completely decoupled, resulting in the localization of $b$ Majorana fermions. %Under the magnetic field, $c$ and $b$ Majorana fermions are hybridized due to the breaking time-reversal symmetry. 
When a magnetic field is applied, these two species of the Majorana fermions begin to hybridize as described in $\mathcal{H}_{bc}^{(1)}$, $\mathcal{H}_{bc}^{(2)}$, and $\mathcal{H}_{bc}^{(3)}$ in Eq.~\eqref{eq:h_eff}.

We clarify the energy band structures of the model defined in the momentum space. Hence, we consider the Fourier transform of the effective Hamiltonian in Eq.~\eqref{eq:h_eff}, which is summarized into
\begin{equation}\label{eq:bloch}
\mathcal{H}_{\mathrm{eff}}=\sideset{}{^{'}}\sum_{\mathbf{k}} \mathbf{a}_{\mathbf{k}}^{\dagger} \mathbf{H}(\mathbf{k})\mathbf{a}_{\mathbf{k}},
\end{equation}
where the summation of the momentum $\mathbf{k}=(k_{x},\ k_{y})$ is taken over half of the first Brillouin zone shown in Fig.~\ref{fig:model}(c). $\mathbf{a}_{\mathbf{k}}=\mathbf{a}^{\dagger}_{-\mathbf{k}}$ is a 16-component vector of the Fourier components of the Majorana fermion operators as $\mathbf{a}_{\mathbf{k}}=(c_{\mathbf{k}, \mathrm{A}},\ \cdots,\ c_{\mathbf{k}, \mathrm{J}},\ b_{\mathbf{k}, \mathrm{A}}^{x},\ b_{\mathbf{k}, \mathrm{C}}^{y},\ b_{\mathbf{k}, \mathrm{E}}^{z},\ b_{\mathbf{k}, \mathrm{F}}^{x},\ b_{\mathbf{k}, \mathrm{H}}^{y}, b_{\mathbf{k}, \mathrm{J}}^{z})$.
%\begin{align}
%\mathbf{a}_{\mathbf{k}}=
%(& c_{\mathbf{k}, \mathrm{A}},\  c_{\mathbf{k}, \mathrm{B}},\  c_{\mathbf{k}, \mathrm{C}},\  c_{\mathbf{k}, \mathrm{D}},\ c_{\mathbf{k}, \mathrm{E}}, \notag\\
%& c_{\mathbf{k}, \mathrm{F}},\ c_{\mathbf{k}, \mathrm{G}},\ c_{\mathbf{k}, \mathrm{H}},\ c_{\mathbf{k}, \mathrm{I}},\ c_{\mathbf{k}, \mathrm{J}}, \notag\\
%& b_{\mathbf{k}, \mathrm{A}}^{x},\ b_{\mathbf{k}, \mathrm{C}}^{y},\ b_{\mathbf{k}, \mathrm{E}}^{z},\ b_{\mathbf{k}, \mathrm{F}}^{x},\ b_{\mathbf{k}, \mathrm{H}}^{y},\ b_{\mathbf{k}, \mathrm{J}}^{z}).
%\end{align}
The Fourier transform of the Majorana fermion operators is defined as
\begin{equation}
a_{\mathbf{r}, \eta}=\sqrt{\frac{2}{N_{\mathrm{c}}}}\sideset{}{^{'}}\sum_{\mathbf{k}}(\mathrm{e}^{\mathrm{i}\mathbf{k}\cdot\mathbf{r}}a_{\mathbf{k}, \eta}+\mathrm{e}^{-\mathrm{i}\mathbf{k}\cdot\mathbf{r}}a_{\mathbf{k}, \eta}^{\dagger}),
\end{equation}
where $a=c$, $b^{x}$, $b^{y}$, or $b^{z}$, and $N_{\mathrm{c}}$ is the number of the unit cells. Each component of the Majorana Bloch Hamiltonian $\mathbf{H}(\mathbf{k})$ is obtained by this Fourier transformation.

\subsection{Chern number}\label{subsec:chern}
We investigate not only the band structures but also the topological nature of the model. In order to characterize the topology, we calculate the Chern number (also known as the TKNN number) which is given by the summation of the Berry curvature
\begin{equation}\label{eq:curvature}
F_{n}(\mathbf{k})=\frac{\partial }{\partial k_{x}}A_{n, y}(\mathbf{k})-\frac{\partial }{\partial k_{y}}A_{n, x}(\mathbf{k}),
\end{equation}
where
\begin{equation}
A_{n, \mu}(\mathbf{k})=-\mathrm{i}\langle u_{n\mathbf{k}}\rvert \frac{\partial}{\partial k_{\mu}}\lvert u_{n\mathbf{k}}\rangle \quad (\mu=x,\ y),
\end{equation}
denotes the Berry connections of $x$ or $y$ components. $\lvert u_{n\mathbf{k}}\rangle$ represents the Bloch wave function of band $n$ with momentum $\mathbf{k}$ of the Majorana Bloch Hamiltonian with the energy eigenvalue $\varepsilon_{n}(\mathbf{k})$:
\begin{equation}
\mathbf{H}(\mathbf{k})\lvert u_{n\mathbf{k}}\rangle = \varepsilon_{n}(\mathbf{k})\lvert u_{n\mathbf{k}}\rangle.
\end{equation}
The band Chern number of the band labeled as $n$ is defined as 
\begin{equation}\label{eq:chern_band}
C_{n}=\frac{1}{2\pi}\int_{\mathbf{k}\in\mathrm{BZ}}\mathrm{d}^{2}k\ F_{n}(\mathbf{k}),
\end{equation}
where the integration is taken over the first Brillouin zone. The total Chern number of the system is given by the summation of the band Chern number for the occupied bands:
\begin{equation}\label{eq:chern_tot}
C=\sum_{n}^{\varepsilon_{n}(\mathbf{k})<0}C_{n}.
\end{equation}
If the Majorana fermion system is in a gapped insulating state, $C$ becomes an integer and its nonzero quantized value indicates that the system is topologically nontrivial. If the Majorana Fermi surface emerges, $C$ becomes ill-defined. 

The band Chern number is defined by the integration in the momentum space in Eq.~\eqref{eq:chern_band}. However, in the actual calculations, we adopt the Fukui-Hatsugai-Suzuki method which is more suitable for numerical calculation~\cite{Fukui2005}. We divide the first Brillouin zone into $10^{6}$ parallelograms with a mesh of $1000$ grid points in each direction of the two reciprocal primitive vectors corresponding $\mathbf{a}_{1}$ and $\mathbf{a}_{2}$ shown in Fig.~\ref{fig:model}(b).

\subsection{Thermal Hall conductivity}\label{subsec:kappa}
The thermal Hall effect is one of the phenomena reflecting the topology of the Majorana fermions.
Here we calculate the bulk thermal Hall conductivity $\kappa_{xy}$ defined as~\cite{Vafek2001, Sumiyoshi2013, Go2019},
\begin{equation}\label{eq:kappa1}
\frac{\kappa_{xy}}{T}=-\frac{1}{2VT^{2}}\int_{-\infty}^{\infty}\mathrm{d}\varepsilon\ \varepsilon^{2}\frac{\partial f(\varepsilon, T)}{\partial \varepsilon}\sum_{\mathbf{k}, n}^{\varepsilon_{n}(\mathbf{k})<\varepsilon}F_{n}(\mathbf{k}),
\end{equation}
where $T$, $V$, and $f(\varepsilon, T)=1/(\mathrm{e}^{\varepsilon/T}+1)$ represent temperature, volume of the system, and Fermi distribution function, respectively. $F_{n}(\mathbf{k})$ is the Berry curvature defined in Eq.~\eqref{eq:curvature}. Changing the integral variable from $\varepsilon$ to $\omega=\varepsilon/T$, we can express Eq.~\eqref{eq:kappa1} as
\begin{equation}\label{eq:kappa2}
\frac{\kappa_{xy}}{T}=\frac{1}{8\pi}\int_{-\infty}^{\infty}\mathrm{d}\omega\ \frac{\omega^{2}}{1+\mathrm{cosh}(\omega)}\tilde{C}(T\omega),
\end{equation}
where
\begin{equation}\label{eq:chern_tilde}
\tilde{C}(T\omega)=\frac{1}{2\pi}\sum_{n}^{\varepsilon_{n}(\mathbf{k})<T\omega}\int_{\mathbf{k}\in\mathrm{BZ}; \varepsilon_{n}(\mathbf{k})<T\omega}\mathrm{d}^{2}k\ F_{n}(\mathbf{k}).
\end{equation}
Here, $\tilde{C}(T\omega)$ is given by a summation of the Berry curvature integrated within the Brillouin zone in the energy range of 
$\varepsilon_{n}(\mathbf{k})<T\omega$. We can calculate this quantity accurately by using the Fukui-Hatsugai-Suzuki method in the same manner as the band and total Chern numbers in Eqs.~\eqref{eq:chern_band} and \eqref{eq:chern_tot}, respectively.
In the zero temperature limit $T\to0$, $\tilde{C}(T\omega)$ becomes the total Chern number $\tilde{C}(0)=C$ and temperature dependence in the right-hand side of Eq.~\eqref{eq:kappa2} vanishes. Then, we obtain
\begin{equation}\label{eq:zero_temp}
\frac{\kappa_{xy}}{T}\to\frac{\pi}{6}\cdot \frac{C}{2},
\end{equation}
by performing the Sommerfeld expansion.
The factor of $1/2$ reflects half-quantization, originating from the fact that each Majorana fermion carries half the degrees of freedom compared to a conventional Dirac fermion.

\section{Result}\label{sec:result}
\subsection{Majorana band structures}\label{subsec:band}
\begin{figure}
    \centering
     \includegraphics[width=\columnwidth,clip]{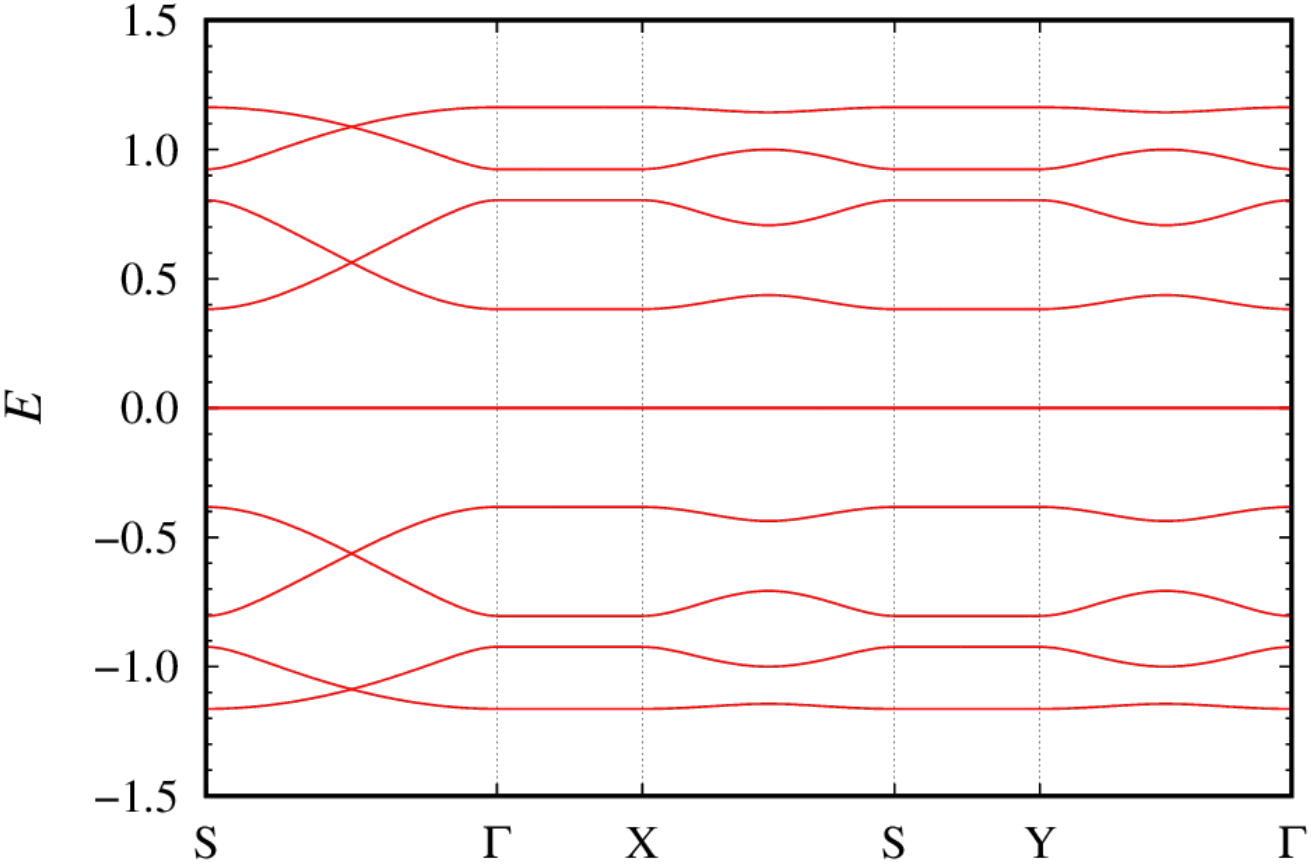}
    \caption{Majorana band structure in the isotropic case at zero magnetic field, shown along the path indicated by the thick red line in Fig.~\ref{fig:model}(c).
}
    \label{fig:band1}
\end{figure}

\begin{figure*}
    \centering
     \includegraphics[width=\linewidth,clip]{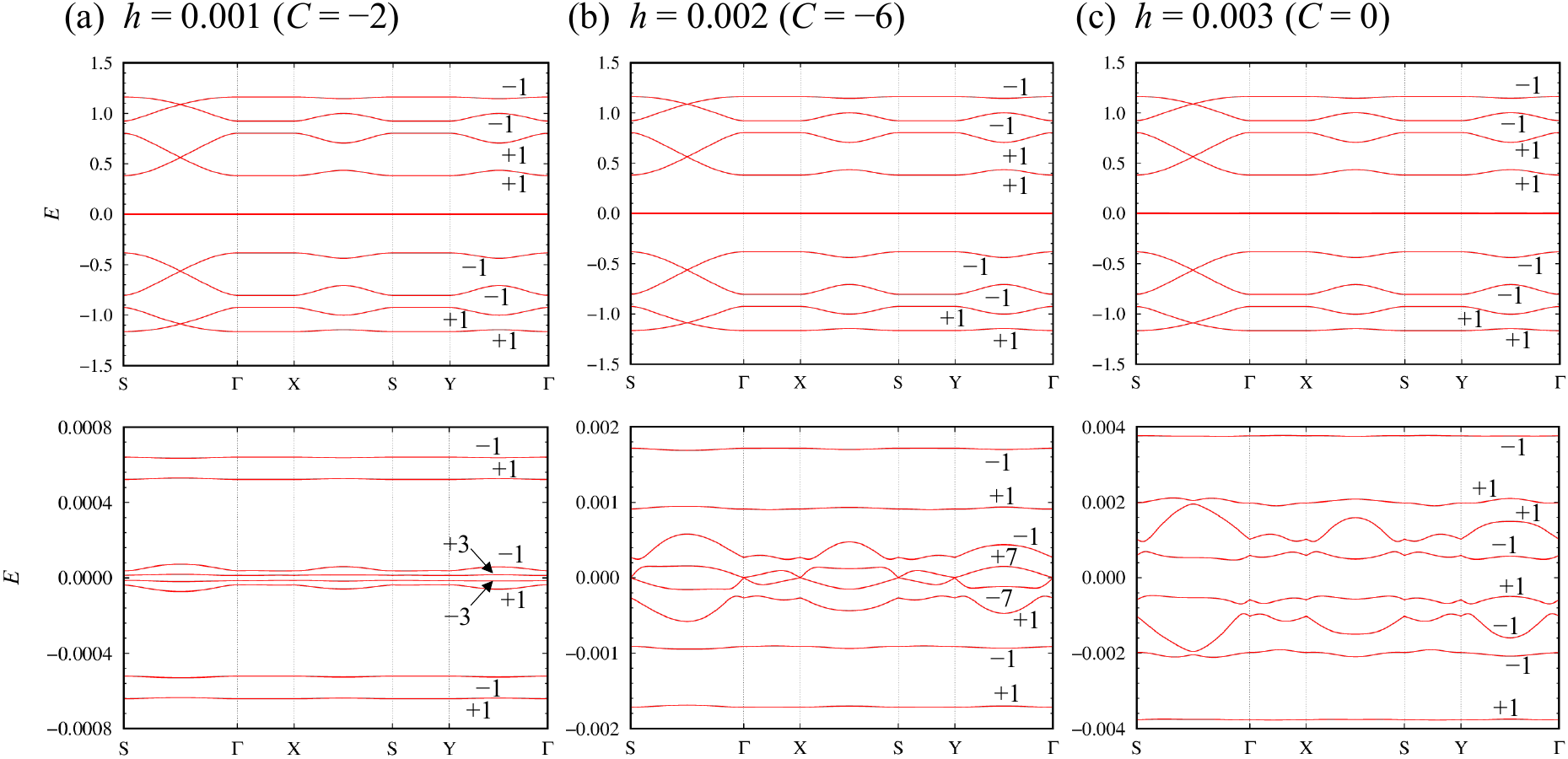}
    \caption{Majorana band structure in the isotropic case at (a) $h=0.001$, (b) $h=0.002$, and (c) $h=0.003$. The lower panels show an enlarged view of the nearly flat bands at low energy. Integers indicate the Chern number of each band, and $C$ indicates the total Chern number which is the sum of the Chern number of the occupied bands.
}
    \label{fig:band2}
\end{figure*}

We first calculate the energy dispersion of the Majorana fermions at zero magnetic field ($\mathbf{h=0}$) by diagonalizing Eq.~\eqref{eq:bloch}. The obtained band structure is shown in Fig.~\ref{fig:band1}. Since the anisotropic case does not differ qualitatively a lot from the isotropic case, we present the result for the isotropic case hereafter in this paper except when discussing anisotropy. Since the $c$ and $b$ Majorana fermions do not hybridize at $\mathbf{h=0}$, ten and six bands of $c$ and $b$ Majorana fermions independently appear, respectively. $b$ Majorana fermions are completely localized at $\mathbf{h=0}$ and therefore the six bands of them are flat and degenerated at zero energy. In addition, two of the ten $c$ Majorana fermion bands are flat bands lying at zero energy, as in the electron system on this lattice~\cite{Hu2023, Wu2024, Chen2024, Ikegami2024}, despite differences in the other dispersive bands due to the presence of the fluxes in the present model.

We next discuss the band dispersion of Majorana fermions in a magnetic field also by diagonalizing Eq.~\eqref{eq:bloch}. In the following, we consider the magnetic field along the $[111]$ direction as $h_x=h_y=h_z=h/\sqrt{3}$. The Majorana bands for magnetic fields of $h=0.001$, $h=0.002$, and $h=0.003$ are shown in Fig.~\ref{fig:band2}.
As mentioned in Sec.~\ref{subsec:perturbation}, the presence of the magnetic field leads to not only second-neighbor hopping processes of $c$ and $b$ Majorana fermions but also hybridization between them, due to broken time-reversal symmetry. As a result, the $b$ Majorana fermions are no longer localized i.e. they become dispersive. As can be seen from the lower panel of the figure, the broken time-reversal symmetry also lifts the degeneracy of the Majorana flat bands at zero magnetic field and opens gaps between them. Hence, we can define the Chern number and band Chern number introduced in Sec.~\ref{subsec:chern} and their values are shown in the figure. 
Each band has a nonzero band Chern number, and in particular, the lowest energy bands exhibit band Chern numbers whose absolute value is larger than 1: $C_{n}= \pm3$ at $h=0.001$ and $C_{n}=\pm 7$ at $h=0.002$. The Majorana flat bands at zero magnetic field are promoted to topological Majorana flat bands in the magnetic field, and the lowest energy ones can show large band Chern numbers. Due to these large band Chern numbers, the total Chern number becomes finite and also larger than 1, which is the case for the honeycomb lattice. 

Here let us note about the symmetries that must be satisfied by the energy bands of the Majorana fermions. We consider three types of symmetries of the Majorana Bloch Hamiltonian in Eq.~\eqref{eq:bloch} below:
\begin{align}
\text{inversion symmetry}\quad P^{-1}\mathbf{H}(-\mathbf{k})P&=\mathbf{H}(\mathbf{k}), \label{eq:inversion}\\
\text{time-reversal symmetry}\quad \Theta^{-1}\mathbf{H}(-\mathbf{k})\Theta&=\mathbf{H}(\mathbf{k}), \label{eq:time-reversal}\\
\text{particle-hole symmetry}\quad \Xi^{-1}\mathbf{H}(-\mathbf{k})\Xi&=-\mathbf{H}(\mathbf{k}). \label{eq:particle-hole}
\end{align}
The inversion symmetry represented by Eq.~\eqref{eq:inversion} is already broken by the lattice geometry shown in Fig.~\ref{fig:model}. At zero magnetic field, the time-reversal and particle-hole symmetries in Eqs.~\eqref{eq:time-reversal} and \eqref{eq:particle-hole}, respectively, are satisfied. The former leads $\varepsilon_{n}(\mathbf{k})=\varepsilon_{n}(-\mathbf{k})$ while the latter leads $\varepsilon_{n}(\mathbf{k})=-\varepsilon_{\bar{n}}(-\mathbf{k})$, where $\bar{n}$ means an index of the band corresponding to the particle-hole transformation of the band with index $n$, for example, $\bar{n}=17-n$ if we assign band indices as $\varepsilon_{1}(\mathbf{k}) \leq \varepsilon_{2}(\mathbf{k}) \leq \cdots \leq \varepsilon_{15}(\mathbf{k}) \leq \varepsilon_{16}(\mathbf{k})$. As a result, we obtain $\varepsilon_{n}(\mathbf{k})=\varepsilon_{n}(-\mathbf{k})=-\varepsilon_{\bar{n}}(-\mathbf{k})=-\varepsilon_{\bar{n}}(\mathbf{k})$, as we can see in Fig.~\ref{fig:band1}. In contrast, in the case of the broken time-reversal symmetry in the magnetic field, only particle-hole symmetry is preserved. Hence, the above relation is no longer valid, and $\varepsilon_{n}(\mathbf{k})=-\varepsilon_{\bar{n}}(-\mathbf{k})\neq -\varepsilon_{\bar{n}}(\mathbf{k})=\varepsilon_{n}(-\mathbf{k})$ is lead instead. In other words, the broken inversion and time-reversal symmetries make Majorana fermions nonreciprocal. We can also see this nonreciprocity in Fig.~\ref{fig:band2}.

\subsection{Topological phase diagram}\label{subsec:phase}
\begin{figure}
    \centering
     \includegraphics[width=\columnwidth,clip]{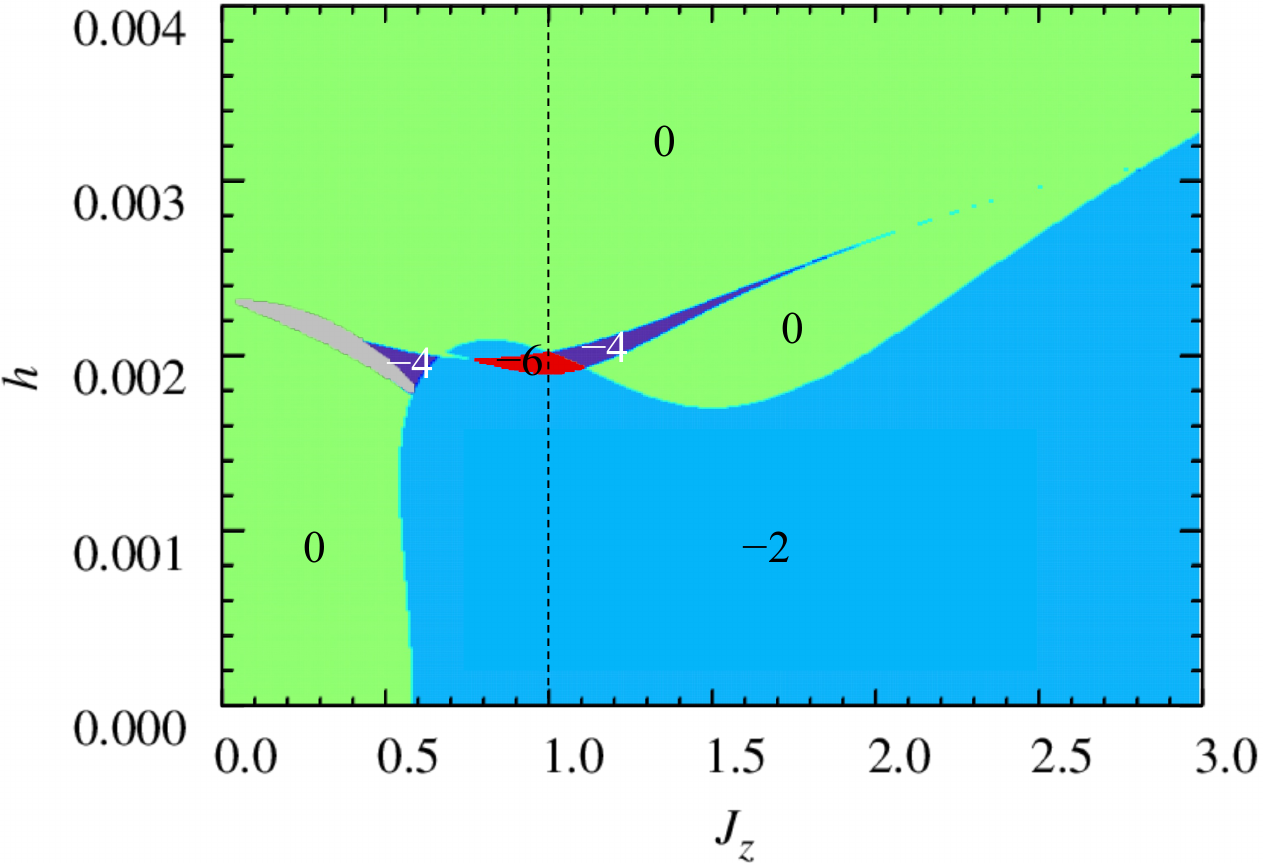}
    \caption{Topological phase diagram of the model. Here we set $J_{x}=J_{y}=(3.0-J_{z})/2$, and $J_{z}=1.0$ corresponds to the isotropic case. The integers denote the total Chern number of each topological phase. Within the gray region, the Majorana Fermi surface emerges, and we cannot define the total Chern number. The dashed line indicates the isotropic case.
}
    \label{fig:phase}
\end{figure}

A topological phase diagram is shown in Fig.~\ref{fig:phase} according to the Chern number calculated by summing up the band Chern numbers of the negative energy bands while varying the anisotropy $J_{z}$ and the magnitude of the magnetic field $h$. The isotropic case discussed in the previous section corresponds to $J_{z}=1.0$. While we find only topological phases with $C=0$, $-2$, and $-6$ in the isotropic case, we observe those with $C=-4$ in the anisotropic regime. 
The topological phase with $C=-2$ extends over a wide region in the low-field regime except for $J_{z}\lesssim 0.6$, while the $C=-4$ phase appears in a narrow region close to the $C=-6$ phase. 
This is in contrast to the previous study on the Haldane model defined on the same lattice, which is an electron analog of our model, topological phases with $C=\pm2$ were found only in narrow parameter regions, and those with $\lvert C \vert >2$ did not appear~\cite{Ikegami2024}. In the gray region for $J_z \lesssim 0.6$ next to the $C=-4$ phase, the Majorana fermion system becomes metallic and the Majorana Fermi surfaces appear.
We note that these diverse topological phases appear in the magnetic field region lower than the flux gap $\Delta \simeq 0.0033$, where the perturbation theory is expected to be justified.

\subsection{Thermal Hall effect}\label{subsec:Hall}
%\subsection{Thermal Hall conductivity}
\begin{figure}
    \centering
     \includegraphics[width=\columnwidth,clip]{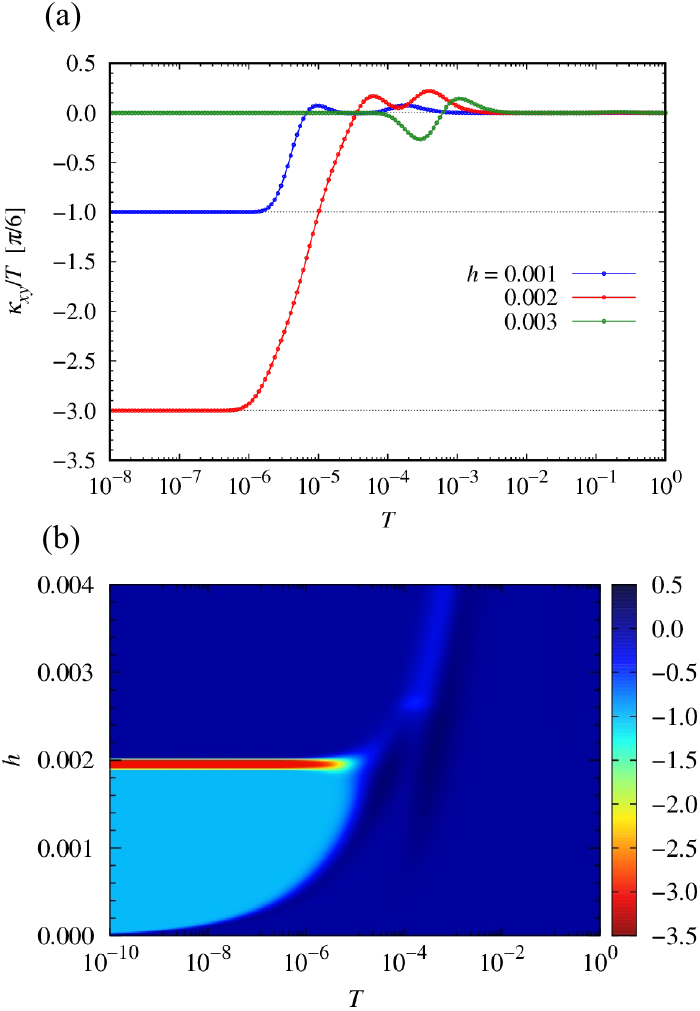}
    \caption{(a) Temperature dependence of the bulk thermal Hall conductivity $\kappa_{xy}$ at the same values of the magnetic field as in Fig.~\ref{fig:band2}. (b) $h$-$T$ plot of $\kappa_{xy}$. We show the value of $\kappa_{xy}$ divided by the temperature $T$ in the unit of $\pi/6$.
}
    \label{fig:kappa}
\end{figure}

In this section, we consider the thermal Hall effect reflecting the large Chern number of the topological phases found in the previous section. 
We calculate the thermal Hall conductivity numerically based on Eq.~\eqref{eq:kappa2} and show its temperature and magnetic field dependence in Fig.~\ref{fig:kappa}. We can confirm the half-quantization of $\kappa_{xy}$ at the low-temperature region $T\lesssim 10^{-6}$, corresponding to zero temperature limit in Eq.~\eqref{eq:zero_temp}.
Note that this temperature scale is much smaller than the flux gap $\Delta \simeq 0.0033$.
 As all topological phases appear in Fig.~\ref{fig:phase} have even Chern number, here $\kappa_{xy}/T$ convergents corresponding integer values in the unit of $\pi/6$. 

In addition, $\kappa_{xy}$ shows oscillating behavior before quantization even in the $C=0$ case. The temperature at which these oscillations appear corresponds to the energy range of the Majorana bands.
These oscillations can be attributed to thermally-excited Majorana fermions, which occupy the bands and contribute additional Berry curvature to Eq.~\eqref{eq:chern_tilde}.
Even in the topologically trivial phases with $C=0$, such oscillations remain observable up to the high-field regime, because the band Chern number remains nonzero at high magnetic fields, as shown in Fig.~\ref{fig:band2}. 
It is worth noting that these oscillations as well as the quantization appear at temperature scale lower than the flux gap $\Delta \simeq 0.0033$, where the perturbation theory is expected to be valid.

\section{Discussion}\label{sec:discussion}
\begin{figure*}
    \centering
     \includegraphics[width=\linewidth,clip]{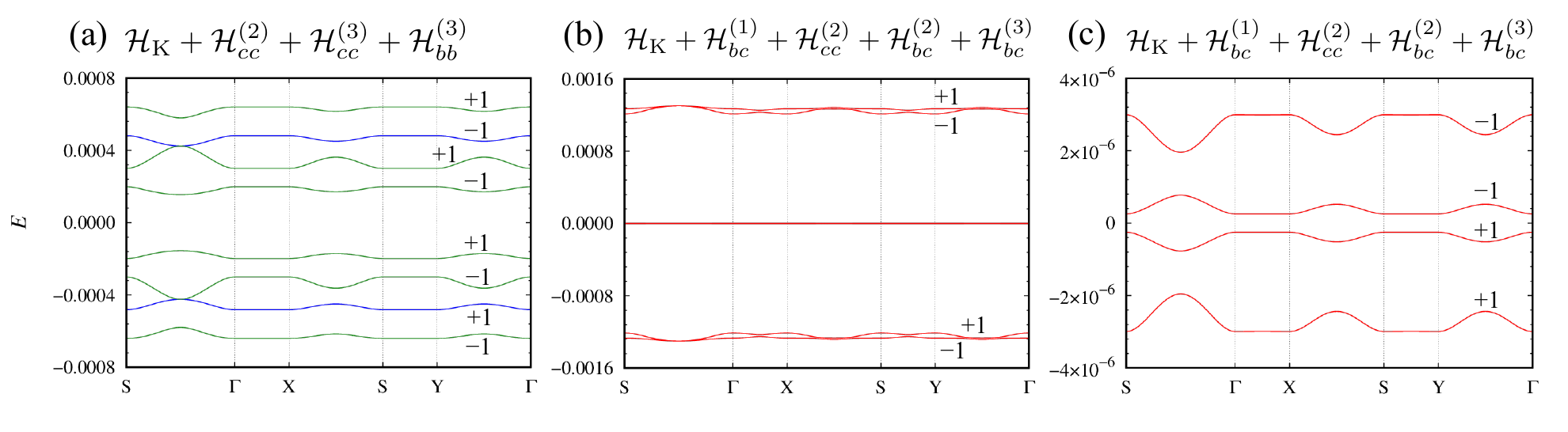}
    \caption{Enlarged views of the nearly flat Majorana bands at low energy and band Chern number at $h=0.002$ calculated without (a) $b$-$c$ hybridization terms, and (b)(c) the second-neighbor $c$-$c$ and $b$-$b$ hopping terms. In (a), bands of $b$ and $c$ Majorana fermions are shown in green and blue, respectively. (c) is an enlarged view around zero energy of (b). The notations are common to those in Fig.~\ref{fig:band2}.
}
    \label{fig:band3}
\end{figure*}

Here we delve into the mechanism of the realization of the topological phases with large Chern numbers. As derived in Sec.~\ref{subsec:perturbation}, there are two important contributions in the perturbation up to the third order: the hybridization of $c$ and $b$ Majorana fermions and the second-neighbor hoppings within $c$ and $b$ Majorana fermions themselves, which also appears in the honeycomb case. Note that $\mathcal{H}^{(2)}_{cc}$ only gives a modulation to the Kitaev couplings $J_{\mu}$ in the unperturbed Hamiltonian in Eq.~\eqref{eq:model}. To clarify which contribution plays a crucial role in realizing large Chern numbers, we first calculate Majorana energy bands and their Chern number by removing the $b\mathchar`- c$ hybridization terms $\mathcal{H}^{(1)}_{bc}$, $\mathcal{H}^{(2)}_{bc}$, and $\mathcal{H}^{(3)}_{bc}$ from the effective Hamiltonian in Eq.~\eqref{eq:h_eff}, and show the result in Fig.~\ref{fig:band3}(a). The figure shows only nearly flat bands near zero energy, where hybridization is important. Due to the lack of $b\mathchar`- c$ hybridization, two types of Majorana fermions are completely decoupled. All bands have only the band Chern number $C_{n}=\pm1$, and the total Chern number is zero. Then we perform computation excluding the second-neighbor hopping terms $\mathcal{H}^{(3)}_{cc}+\mathcal{H}^{(3)}_{bb}$, and  show the result in Figs.~\ref{fig:band3}(b) and \ref{fig:band3}(c). Also, in this case, all bands only have $C_{n}=\pm1$, and the total Chern number only reaches at most $\lvert C\rvert=2$. 

From these observations, we can conclude that both the second-neighbor hoppings and $b\mathchar`- c$ hybridization are essential: their cooperation plays a crucial role in the emergence of the topological Majorana flat bands with large band Chern number. These contributions come from (i) the coexistence of the unpaired $b$ and $c$ Majorana fermions, and (ii) $\mathbb{Z}_{2}$ fluxes consisting of the paired $b$ Majorana fermions. It is worth noting that both contributions are direct consequences of fractionalization of quantum spins.
This indicates that our findings of the flat bands with large Chern numbers are unique to fractionalized excitations in quantum spin liquids. Note that an electron analogue does not realize $\lvert C\rvert >2$~\cite{Ikegami2024}, as mentioned in Sec.~\ref{subsec:phase}.

Finally, let us discuss the similarity and diffecence between our results with periodic vacancies and the cases of isolated vacancies.
Effect of isolated or pair of isolated site vacancies in the Kitaev model have been intensively studied~\cite{Trousselet2011, Willans2011, Santhosh2012, Udagawa2018, Nasu2020, Kao2021, Udagawa2021, Kao2021a, Dantas2022, Li2023, Takahashi2023, Kao2024, Kao2024a, Yatsuta2024, Xiao2025}. In the case of isolated vacancies, unpaired $b$ Majorana fermions are also localized around the defect sites, forming zero modes, at zero magnetic field~\cite{Santhosh2012, Udagawa2021,Takahashi2023, Kao2024, Kao2024a,Yatsuta2024}. This zero modes remain even under a magnetic field, indicating nonlocal correlation between them~\cite{Takahashi2023}. 
In contrast, our model introduces site vacancies periodically, forming a vacancy lattice.
This periodic structure enables the $b$ Majorana fermions to acquire dynamics under a magnetic field. Consequently the zero modes are lifted and become topological flat bands with large Chern numbers.

\section{Summary}\label{sec:summary}
To summarize, we have studied the Majorana flat bands and their topology on the Bishamon-kikko lattice in the presence of the magnetic field by using the perturbation theory, and how to detect the large Chern number they have. We found that the Kitaev spin liquid on the Bishamon-kikko lattice hosts the Majorana flat bands at zero magnetic field and topological Majorana flat bands with large band Chern numbers under a magnetic field. Due to the large band Chern numbers, several topological phases with $\lvert C \rvert > 1$ appear depending on the magnitude of the magnetic field and the anisotropy of the exchange interaction. We clarified that the emergence of the topological Majorana flat bands with large band Chern numbers reflects the fractionalization of quantum spins into two types ($c$ and $b$) of Majorana fermions. We also demonstrated that these large total Chern numbers are detectable by measurement of the thermal Hall conductivity, which exhibits oscillations reflecting Berry curvature, even in the topologically trivial phases.

As we mentioned in Sec.~\ref{subsec:latticemodel}, we can consider various types of the Bishamon-kikko lattice with the unit motif shown in Fig.~\ref{fig:model}(a). Other tilings can change not only the symmetries of the lattice but also the band structures of the Majorana fermions and their Chern number. In addition, other ways of site depletion are also intriguing. Changes in the unit motif and unit cell due to different depletion periods can also alter the band structure and Chern number of the Majorana fermions, as well as the flux configuration of the ground state, leading to more diverse topology. Random site depletions in the Kitaev magnets have been investigated experimentally in Na$_{2}$(Ir$_{1-x}$Ti$_{x}$)O$_{3}$~\cite{Manni2014R}, (Na$_{1-x}$Li$_{x}$)$_{2}$IrO$_{3}$~\cite{Cao2013, Manni2014, Rolfs2015, Gupta2016, Hermann2017, Simutis2018}, Li$_{2}$(Ir$_{1-x}$Ti$_{x}$)O$_{3}$~\cite{Manni2014R}, Li$_{2}$Ir$_{1-x}$Ru$_{x}$O$_{3}$~\cite{Lei2014},  $\alpha$-Ru$_{1-x}$Ir$_{x}$Cl$_{3}$~\cite{Lampen-Kelley2017, Do2018, Do2020, Beak2020}, $\alpha$-Ru$_{1-x}$Cr$_{x}$Cl$_{3}$~\cite{Bastien2019}, and Os$_{x}$Cl$_{3}$~\cite{Kataoka2020}. It would be intriguing to develop methods for controlling the configuration of these defects, enabling the exploration of the diverse topological properties of Majorana fermions.

In electron systems, topological flat bands provide a promising and tunable platform to realize a variety of intriguing correlated topological states of matter, such as fractional topological insulators and fractional anomalous Hall effect~\cite{Bergholtz2013, Neupert2015}. 
In addition, flat bands and their topology have garnered increasing attention in recent years due to their strong relationship to the quantum metric~\cite{Peotta2015, Torma2022, Mera2022, Guo2024, Ying2024, Chau2024, Penttila2025}.
Our study sheds light on the topology of the Majorana flat bands and provides an attractive playground for novel correlated topological phenomena of the Majorana fermions. While the present study employed a noninteracting effective model of the Majorana fermions, including many-body interactions may yield richer topological quantum phases.
Furthermore, in the context of quantum spin liquids, our findings clarify an unprecedented scenario for the emergence of the topological Majorana flat bands with large Chern numbers reflecting the fractionalization of quantum spins. They would stimulate the exploration of a new type of Kitaev materials exhibiting rich topology from topological Majorana flat bands.

\begin{acknowledgments}
The authors thank K. Ido, Y. Kato, T. Misawa, J. Nasu, T. Okubo, and S. Ikegami for fruitful discussions. K.F. also thanks K. Shimizu, M. O. Takahashi, and I. Murahashi for constructive suggestions.
Parts of the numerical calculations have been done using the facilities of the Supercomputer Center, the Institute for Solid State Physics, the University of Tokyo, the Information Technology Center, the University of Tokyo, and the Center for Computational Science, University of Tsukuba.
This work was supported by Japan Society for the Promotion of Science (JSPS) KAKENHI Grant Nos. 19H05825, 20H00122, 24K17009, and 25H01247.
\end{acknowledgments}
%\newpage
\appendix
\section{Confirmation of the $\pi$-flux sector}\label{app:flux}
\begin{figure}
    \centering
     \includegraphics[width=\columnwidth,clip]{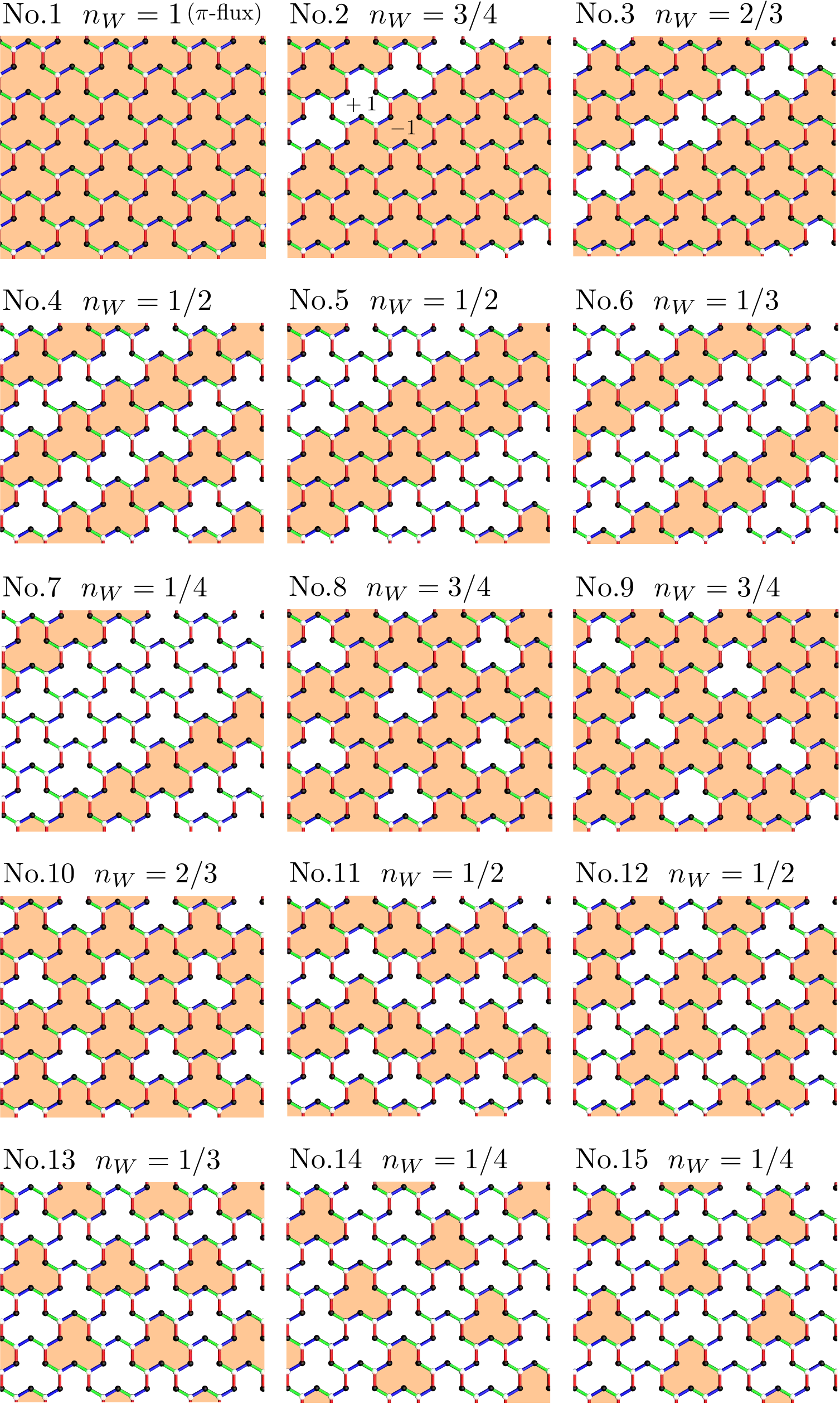}
    \caption{Schematic picture of the 15 flux patterns we compare the ground-state energy. The $\mathbb{Z}_{2}$ gauge flux $W_{p}=-1$ is excited at the plaquettes colored in orange. $n_{W}$ indicates the flux density.
}
    \label{fig:fluxpatterns}
\end{figure}

\begin{figure}
    \centering
     \includegraphics[width=\columnwidth,clip]{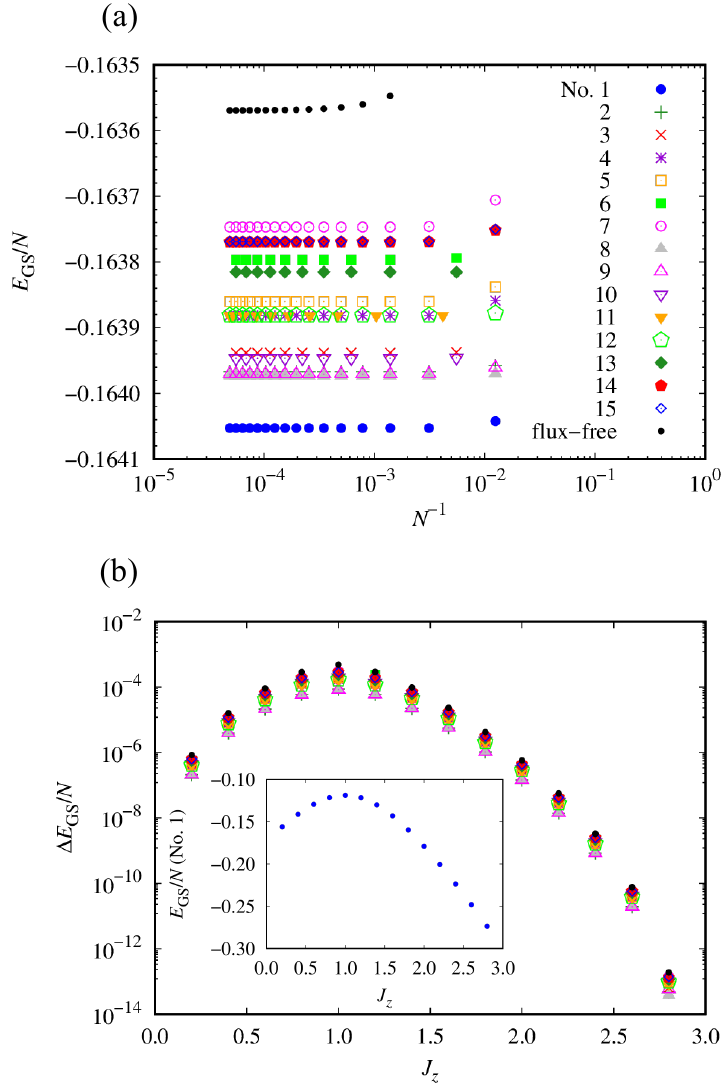}
    \caption{(a) Comparison of $N$ dependence of the ground-state energy per site of each flux pattern shown in Fig.~\ref{fig:fluxpatterns} in addition to the flux-free state. $N$ is the number of lattice sites. (b) Comparison of $J_{z}$ dependence of the ground-state energy per site of each flux pattern calculated with $N=18000$ ($N=19440$ only for No.11). Here we set $J_{x}=J_{y}=(3.0-J_{z})/2$, and show the difference from the ground-state energy of the flux pattern No. 1 ($\pi$-flux state) shown in the inset.
}
    \label{fig:Egs}
\end{figure}

In this appendix, we present the numerical results that confirm that the $\pi$-flux state is realized as the ground state at zero magnetic field. We consider 16 flux configurations: 15 flux patterns with several flux densities shown in Fig.~\ref{fig:fluxpatterns} and the flux-free state ($W_{p}=+1$ for all plaquettes), and compare the ground-state energy. First, we calculate the ground-state energy $E_{\mathrm{GS}}$ of each flux configuration with isotropic Kitaev interactions $J_{x}=J_{y}=J_{z}=1$ by numerically diagonalizing Hamiltonian in Eq.~\eqref{eq:H_K} with a periodic boundary condition. By tuning the values of $u^{\mu}_{ij}$s we realize each flux configuration shown in Fig.~\ref{fig:fluxpatterns}.
We show the system size dependence of the ground-state energy per site in Fig.~\ref{fig:Egs}(a). We can see that the ground-state energy decreases monotonically with increasing flux density, and the $\pi$-flux state has the lowest ground-state energy while the flux-free state has the highest one. 
We also calculate the nearest-neighbor equal-time spin correlation function as $\langle S^{\mu}_{i}S^{\mu}_{j}\rangle\simeq 0.1367$ for the $\pi$-flux state and confirm that this value is consistent with the ground-state energy we obtained here.
Note that the energies of all flux configurations, except for the flux-free state, converge rapildy with increasing the system size, indicating that these states are gapped.

We also confirm that the $\pi$-flux state gives the ground state in the anisotropic case. In Fig.~\ref{fig:Egs}(b), we show $J_{z}$ dependence of the ground-state energy of each flux configuration. We show the data calculated with the cluster containing $N=18000$ ($N=19440$ only for No. 11). Here we change $J_{z}$ and set $J_{x}=J_{y}=(3.0-J_{z})/2$ as in Fig.~\ref{fig:phase}. Since the energy differences are quite small and the $\pi$-flux state gives the lowest energy regardless $J_{z}$, we show the difference of the ground-state energy per site with the $\pi$-flux state: $\Delta E_{\mathrm{GS}}/N=\{E_{\mathrm{GS}}(\text{each flux configuration})-E_{\mathrm{GS}}(\pi\text{-flux state})\}/N$. The data for the $\pi$-flux state are shown in the inset.

\section{Explicit expression of the perturbative corrections}\label{app:perturbation}
In this appendix, we show the explicit expression of the perturbative corrections in our effective Hamiltonian in Eq.~\eqref{eq:h_eff}. As the first-order corrections of order $\mathcal{O}(h_{\mu})$, we obtain onsite $b$-$c$ hybridizations:
\begin{align}
\mathcal{H}_{bc}^{(1)}=-\frac{\mathrm{i}}{2}\sum_{\mathbf{r}}\big[
& h_{x}(b_{\mathbf{r}, \mathrm{A}}^{x}c_{\mathbf{r}, \mathrm{A}} + b_{\mathbf{r},\mathrm{F}}^{x}c_{\mathbf{r},\mathrm{F}}) \notag\\
+& h_{y}( b_{\mathbf{r}, \mathrm{C}}^{y}c_{\mathbf{r}, \mathrm{C}} + b_{\mathbf{r}, \mathrm{H}}^{y}c_{\mathbf{r}, \mathrm{H}}) \notag\\
+& h_{z}( b_{\mathbf{r}, \mathrm{E}}^{z}c_{\mathbf{r}, \mathrm{E}} + b_{\mathbf{r}, \mathrm{J}}^{z}c_{\mathbf{r}, \mathrm{J}})\big].
\end{align}
From the second-order perturbation of order $\mathcal{O}(h_{\mu}^{2}/\Delta)$, we obtain two types of corrections: nearest-neighbor $c$-$c$ and $b$-$c$ hoppings given as
\begin{align}
\mathcal{H}_{cc}^{(2)}=\frac{\mathrm{i}}{2\Delta}\sum_{\mathbf{r}}\big[
 h_{x}^{2}( & u_{\mathrm{CB}}^{x} c_{\mathbf{r},\mathrm{C}}c_{\mathbf{r},\mathrm{B}}+u_{\mathrm{ED}}^{x} c_{\mathbf{r},\mathrm{E}}c_{\mathbf{r},\mathrm{D}} \notag\\
 +& u_{\mathrm{HG}}^{x} c_{\mathbf{r},\mathrm{H}}c_{\mathbf{r},\mathrm{G}} + u_{\mathrm{JI}}^{x} c_{\mathbf{r},\mathrm{J}}c_{\mathbf{r},\mathrm{I}}) \notag\\
+ h_{y}^{2}( & u_{\mathrm{EG}}^{y} c_{\mathbf{r},\mathrm{E}}c_{\mathbf{r},\mathrm{G}}+u_{\mathrm{AD}}^{y} c_{\mathbf{r}+\mathbf{a}_{2},\mathrm{A}}c_{\mathbf{r},\mathrm{D}} \notag\\
+& u_{\mathrm{JB}}^{y} c_{\mathbf{r},\mathrm{J}}c_{\mathbf{r}+\mathbf{a}_{1},\mathrm{B}}+u_{\mathrm{FI}}^{y} c_{\mathbf{r}+\mathbf{a}_{2},\mathrm{F}}c_{\mathbf{r},\mathrm{I}}) \notag\\
+ h_{z}^{2} ( & u_{\mathrm{AB}}^{z} c_{\mathbf{r},\mathrm{A}}c_{\mathbf{r},\mathrm{B}} + u_{\mathrm{CD}}^{z} c_{\mathbf{r},\mathrm{C}}c_{\mathbf{r},\mathrm{D}} \notag\\
& u_{\mathrm{FG}}^{z} c_{\mathbf{r},\mathrm{F}}c_{\mathbf{r},\mathrm{G}}+u_{\mathrm{HI}}^{z} c_{\mathbf{r},\mathrm{H}}c_{\mathbf{r},\mathrm{I}})\big],
\end{align}
and
\begin{align}
\mathcal{H}_{bc}^{(2)}=\frac{\mathrm{i}}{2\Delta}\sum_{\mathbf{r}}\big[
h_{x}h_{y}( & u_{\mathrm{ED}}^{x} b_{\mathbf{r},\mathrm{E}}^{z}c_{\mathbf{r},\mathrm{D}}-u_{\mathrm{EG}}^{y} b_{\mathbf{r},\mathrm{E}}^{z}c_{\mathbf{r},\mathrm{G}} \notag\\
+& u_{\mathrm{JI}}^{x} b_{\mathbf{r},\mathrm{J}}^{z}c_{\mathbf{r},\mathrm{I}}-u_{\mathrm{JB}}^{y} b_{\mathbf{r},\mathrm{J}}^{z}c_{\mathbf{r}+\mathbf{a}_{1},\mathrm{B}}) \notag\\
+h_{y}h_{z}( & u_{\mathrm{AD}}^{y} b_{\mathbf{r}+\mathbf{a}_{2},\mathrm{A}}^{x}c_{\mathbf{r},\mathrm{D}} - u_{\mathrm{AB}}^{z} b_{\mathbf{r},\mathrm{A}}^{x}c_{\mathbf{r},\mathrm{B}} \notag\\
+& u_{\mathrm{FI}}^{y} b_{\mathbf{r}+\mathbf{a}_{2},\mathrm{F}}^{x}c_{\mathbf{r},\mathrm{I}} - u_{\mathrm{FG}}^{z} b_{\mathbf{r},\mathrm{F}}^{x}c_{\mathbf{r},\mathrm{G}}) \notag\\
+h_{z}h_{x}( & u_{\mathrm{CD}}^{z} b_{\mathbf{r},\mathrm{C}}^{y}c_{\mathbf{r},\mathrm{D}} - u_{\mathrm{CB}}^{x} b_{\mathbf{r},\mathrm{C}}^{y}c_{\mathbf{r},\mathrm{B}} \notag\\
+& u_{\mathrm{HI}}^{z} b_{\mathbf{r},\mathrm{H}}^{y}c_{\mathbf{r},\mathrm{I}} - u_{\mathrm{HG}}^{x} b_{\mathbf{r},\mathrm{H}}^{y}c_{\mathbf{r},\mathrm{G}})\big],
\end{align}
respectively [see also Figs.~\ref{fig:hopping}(a) and \ref{fig:hopping}(b)]. Note that $\mathcal{H}^{(2)}_{cc}$ has the same form as the Kitaev model $\mathcal{H}_{\mathrm{K}}$ in Eq.~\eqref{eq:H_K}, and it  just shifts the value of the Kitaev couplings $J_{\mu}$.
As the third-order corrections of order $\mathcal{O}(h_{\mu}^{3}/\Delta^{2})$, there are three types of corrections: second-neighbor $c$-$c$, $b$-$c$, and $b$-$b$ hoppings given as
\begin{align}
\mathcal{H}_{cc}^{(3)}=\mathrm{i}&\frac{3 h_{x}h_{y}h_{z}}{4\Delta^{2}}\sum_{\mathbf{r}} \notag\\
\times\big[
& u_{\mathrm{JB}}^{y}u_{\mathrm{AB}}^{z}c_{\mathbf{r}-\mathbf{a}_{1},\mathrm{J}}c_{\mathbf{r},\mathrm{A}} + u_{\mathrm{AD}}^{y}u_{\mathrm{CD}}^{z}c_{\mathbf{r}+\mathbf{a}_{2},\mathrm{A}}c_{\mathbf{r},\mathrm{C}} \notag\\
+& u_{\mathrm{AB}}^{z}u_{\mathrm{CB}}^{x}c_{\mathbf{r},\mathrm{A}}c_{\mathbf{r},\mathrm{C}} + u_{\mathrm{CD}}^{z}u_{\mathrm{ED}}^{x}c_{\mathbf{r},\mathrm{C}}c_{\mathbf{r},\mathrm{E}} \notag\\
+& u_{\mathrm{CB}}^{x}u_{\mathrm{JB}}^{y}c_{\mathbf{r},\mathrm{C}}c_{\mathbf{r}-\mathbf{a}_{1},\mathrm{J}} + u_{\mathrm{ED}}^{x}u_{\mathrm{AD}}^{y}c_{\mathbf{r},\mathrm{E}}c_{\mathbf{r}+\mathbf{a}_{2},\mathrm{A}} \notag\\
+& u_{\mathrm{EG}}^{y}u_{\mathrm{FG}}^{z}c_{\mathbf{r},\mathrm{E}}c_{\mathbf{r},\mathrm{F}}
+ u_{\mathrm{FI}}^{y}u_{\mathrm{HI}}^{z}c_{\mathbf{r}+\mathbf{a}_{2},\mathrm{F}}c_{\mathbf{r},\mathrm{H}} \notag\\
+& u_{\mathrm{FG}}^{z}u_{\mathrm{HG}}^{x}c_{\mathbf{r},\mathrm{F}}c_{\mathbf{r},\mathrm{H}} + u_{\mathrm{HI}}^{z}u_{\mathrm{JI}}^{x}c_{\mathbf{r},\mathrm{H}}c_{\mathbf{r},\mathrm{J}} \notag\\
+& u_{\mathrm{HG}}^{x}u_{\mathrm{EG}}^{y}c_{\mathbf{r},\mathrm{H}}c_{\mathbf{r},\mathrm{E}} + u_{\mathrm{JI}}^{x}u_{\mathrm{FI}}^{y}c_{\mathbf{r},\mathrm{J}}c_{\mathbf{r}+\mathbf{a}_{2},\mathrm{F}}\big],
\end{align}

\begin{align}
\mathcal{H}_{bc}^{(3)}=\mathrm{i}&\frac{3}{4\Delta^{2}}\sum_{\mathbf{r}} \notag\\
\times\big[h_{x}^{2}h_{y}(& u_{\mathrm{AD}}^{y}u_{\mathrm{CD}}^{z}c_{\mathbf{r}+\mathbf{a}_{2},\mathrm{A}}b_{\mathbf{r},\mathrm{C}}^{y} + u_{\mathrm{CD}}^{z}u_{\mathrm{ED}}^{x}b_{\mathbf{r},\mathrm{C}}^{y}c_{\mathbf{r},\mathrm{E}} \notag\\
+& u_{\mathrm{FI}}^{y}u_{\mathrm{HI}}^{z}c_{\mathbf{r}+\mathbf{a}_{2},\mathrm{F}}b_{\mathbf{r},\mathrm{H}}^{y} + u_{\mathrm{HI}}^{z}u_{\mathrm{JI}}^{x}b_{\mathbf{r},\mathrm{H}}^{y}c_{\mathbf{r},\mathrm{J}}) \notag\\
+h_{x}h_{y}^{2}(& u_{\mathrm{AB}}^{z}u_{\mathrm{CB}}^{x}c_{\mathbf{r},\mathrm{C}}b_{\mathbf{r},\mathrm{A}}^{x} + u_{\mathrm{JB}}^{y}u_{\mathrm{AB}}^{z}b_{\mathbf{r},\mathrm{A}}^{x}c_{\mathbf{r}-\mathbf{a}_{1},\mathrm{J}} \notag\\
+& u_{\mathrm{FG}}^{z}u_{\mathrm{HG}}^{x}c_{\mathbf{r},\mathrm{H}}b_{\mathbf{r},\mathrm{F}}^{x} + u_{\mathrm{EG}}^{y}u_{\mathrm{FG}}^{z}b_{\mathbf{r},\mathrm{F}}^{x}c_{\mathbf{r},\mathrm{E}}) \notag\\
+h_{y}^{2}h_{z}(& u_{\mathrm{CD}}^{z}u_{\mathrm{ED}}^{x}c_{\mathbf{r},\mathrm{C}}b_{\mathbf{r},\mathrm{E}}^{z} + u_{\mathrm{ED}}^{x}u_{\mathrm{AD}}^{y}b_{\mathbf{r},\mathrm{E}}^{z}c_{\mathbf{r}+\mathbf{a}_{2},\mathrm{A}} \notag\\
+& u_{\mathrm{HI}}^{z}u_{\mathrm{JI}}^{x}c_{\mathbf{r},\mathrm{H}}b_{\mathbf{r},\mathrm{J}}^{z} + u_{\mathrm{JI}}^{x}u_{\mathrm{FI}}^{y}b_{\mathbf{r},\mathrm{J}}^{z}c_{\mathbf{r}+\mathbf{a}_{2},\mathrm{F}}) \notag\\
+h_{y}h_{z}^{2}(& u_{\mathrm{CB}}^{x}u_{\mathrm{JB}}^{y}c_{\mathbf{r}-\mathbf{a}_{1},\mathrm{J}}b_{\mathbf{r},\mathrm{C}}^{y} + u_{\mathrm{AB}}^{z}u_{\mathrm{CB}}^{x}b_{\mathbf{r},\mathrm{C}}^{y}c_{\mathbf{r},\mathrm{A}} \notag\\
+& u_{\mathrm{HG}}^{x}u_{\mathrm{EG}}^{y}c_{\mathbf{r},\mathrm{E}}b_{\mathbf{r},\mathrm{H}}^{y} + u_{\mathrm{FG}}^{z}u_{\mathrm{HG}}^{x}b_{\mathbf{r},\mathrm{H}}^{y}c_{\mathbf{r},\mathrm{F}}) \notag\\
+h_{z}^{2}h_{x}(& u_{\mathrm{ED}}^{x}u_{\mathrm{AD}}^{y}c_{\mathbf{r},\mathrm{E}}b_{\mathbf{r}+\mathbf{a}_{2},\mathrm{A}}^{x} + u_{\mathrm{AD}}^{y}u_{\mathrm{CD}}^{z}b_{\mathbf{r}+\mathbf{a}_{2},\mathrm{A}}^{x}c_{\mathbf{r},\mathrm{C}} \notag\\
+& u_{\mathrm{JI}}^{x}u_{\mathrm{FI}}^{y}c_{\mathbf{r},\mathrm{J}}b_{\mathbf{r}+\mathbf{a}_{2},\mathrm{F}}^{x} + u_{\mathrm{FI}}^{y}u_{\mathrm{HI}}^{z}b_{\mathbf{r}+\mathbf{a}_{2},\mathrm{F}}^{x}c_{\mathbf{r},\mathrm{H}}) \notag\\
+h_{z}h_{x}^2(& u_{\mathrm{JB}}^{y}u_{\mathrm{AB}}^{z}c_{\mathbf{r},\mathrm{A}}b_{\mathbf{r}-\mathbf{a}_{1},\mathrm{J}}^{z} + u_{\mathrm{CB}}^{x}u_{\mathrm{JB}}^{y}b_{\mathbf{r}-\mathbf{a}_{1},\mathrm{J}}^{z}c_{\mathbf{r},\mathrm{C}} \notag\\
+& u_{\mathrm{EG}}^{y}u_{\mathrm{FG}}^{z}c_{\mathbf{r},\mathrm{F}}b_{\mathbf{r},\mathrm{E}}^{z} + u_{\mathrm{HG}}^{x}u_{\mathrm{EG}}^{y}b_{\mathbf{r},\mathrm{E}}^{z}c_{\mathbf{r},\mathrm{H}})
\big],
\end{align}
and
\begin{align}
\mathcal{H}_{bb}^{(3)}=\mathrm{i}&\frac{3}{4\Delta^{2}}\sum_{\mathbf{r}} \notag\\
\times\big[& h_{x}^{2}h_{y}(u_{\mathrm{JB}}^{y}u_{\mathrm{AB}}^{z}b_{\mathbf{r}-\mathbf{a}_{1},\mathrm{J}}^{z}b_{\mathbf{r},\mathrm{A}}^{x} + u_{\mathrm{EG}}^{y}u_{\mathrm{FG}}^{z}b_{\mathbf{r},\mathrm{E}}^{z}b_{\mathbf{r},\mathrm{F}}^{x}) \notag\\
+& h_{x}h_{y}^{2}(u_{\mathrm{CD}}^{z}u_{\mathrm{ED}}^{x}b_{\mathbf{r},\mathrm{C}}^{y}b_{\mathbf{r},\mathrm{E}}^{z} + u_{\mathrm{HI}}^{z}u_{\mathrm{JI}}^{x}b_{\mathbf{r},\mathrm{H}}^{y}b_{\mathbf{r},\mathrm{J}}^{z}) \notag\\
+& h_{y}^{2}h_{z}(u_{\mathrm{AB}}^{z}u_{\mathrm{CB}}^{x}b_{\mathbf{r},\mathrm{A}}^{x}b_{\mathbf{r},\mathrm{C}}^{y} + u_{\mathrm{FG}}^{z}u_{\mathrm{HG}}^{x}b_{\mathbf{r},\mathrm{F}}^{x}b_{\mathbf{r},\mathrm{H}}^{y}) \notag\\
+& h_{y}h_{z}^{2}(u_{\mathrm{ED}}^{x}u_{\mathrm{AD}}^{y}b_{\mathbf{r},\mathrm{E}}^{z}b_{\mathbf{r}+\mathbf{a}_{2},\mathrm{A}}^{x} + u_{\mathrm{JI}}^{x}u_{\mathrm{FI}}^{y}b_{\mathbf{r},\mathrm{J}}^{z}b_{\mathbf{r}+\mathbf{a}_{2},\mathrm{F}}^{x}) \notag\\
+& h_{z}^{2}h_{x}(u_{\mathrm{CB}}^{x}u_{\mathrm{JB}}^{y}b_{\mathbf{r},\mathrm{C}}^{y}b_{\mathbf{r}-\mathbf{a}_{1},\mathrm{J}}^{z} + u_{\mathrm{HG}}^{x}u_{\mathrm{EG}}^{y}b_{\mathbf{r},\mathrm{H}}^{y}b_{\mathbf{r},\mathrm{E}}^{z}) \notag\\
+& h_{z}h_{x}^{2}(u_{\mathrm{AD}}^{y}u_{\mathrm{CD}}^{z}b_{\mathbf{r}+\mathbf{a}_{2},\mathrm{A}}^{x}b_{\mathbf{r},\mathrm{C}}^{y} + u_{\mathrm{FI}}^{y}u_{\mathrm{HI}}^{z}b_{\mathbf{r}+\mathbf{a}_{2},\mathrm{F}}^{x}b_{\mathbf{r},\mathrm{H}}^{y})\big],
\end{align}
respectively. The directions of each hopping process are shown in Figs.~\ref{fig:hopping}(c) and \ref{fig:hopping}(d).

\bibliography{library}

\end{document}